\documentclass[aps,showpacs]{revtex4}
\usepackage{bm}
\usepackage{graphicx}

\newcommand{\beq}{\begin{equation}}
\newcommand{\eeq}{\end{equation}}
\newcommand{\bd}[1]{ \mbox{\boldmath $#1$}}
\begin{document}
\def\ii{\'\i}
\title{A schematic model for QCD. III: \\
Hadronic states.}
\author{
M. Nu\~nez V.,
S. Lerma H. and
P. O. Hess,}
\affiliation{Inst. Cs Nucleares, Univ.
Nacional Aut\'onoma de M\'exico,
Apdo. Postal 70-543, M\'exico 04510 D.F.}
\author{S. Jesgarz}
\affiliation{Inst. Fis., Univ. de S\~ao Paulo,
CP 66318, S\~ao Paulo, 05315-970, SP, Brasil}
\author{O. Civitarese and
M. Reboiro}
\affiliation{Dep. Fis., Univ. Nac. de La Plata,
c.c. 67 1900, La Plata, Argentina. }

\begin{abstract}
{The hadronic spectrum  obtained  in the framework of
a QCD-inspired schematic model, is presented. The
model is the extension of a previous version, whose basic degrees of
freedom are constituent quarks and antiquarks, and gluons. The
interaction between quarks and gluons is a phenomenological
interaction and its parameters are fixed from data. The classification
of the states, in terms of quark and antiquark and gluon
configurations is based on symmetry considerations, and it is
independent of the chosen interaction. Following this procedure,
nucleon and $\Delta$ resonances are identified, as well as various
penta- and hepta-quarks states. The lowest pentaquarks state is
predicted at 1.5 GeV and it has negative parity, while the lowest
hepta-quarks state has positive parity and its energy is of the
order of 2.5 GeV. } 
\end{abstract}
\maketitle

\section{Introduction}

In previous papers \cite{paperI,paperII,paperIII}, we have proposed a schematic
model aimed at the description of the non-perturbative regime of
QCD.
The main concepts about the model can be found in \cite{paperI}, together with the
applications to the meson spectrum of QCD. Along this line,
Ref. \cite{paperII} deals
with the appearance of phase transitions and condensates.
A preliminary set of results, about the energy and parity of
systems of the type $q^3(q{\bar q})$ (pentaquarks), and
$q^3(q{\bar q})^2$ (heptaquarks), was presented in
\cite{paperIII}.

As it was discussed in \cite{paperI}, this model
describes reasonable well the main features of the meson spectrum
of QCD. This is a nice result since it shows that, also in the
conditions of low energy QCD, e.g.; large coupling constants and
non-conservation of particle number, the use of simple models may
be very useful. The model of \cite{paperI} belongs to the class of
models \cite{lipkin} which may be solved by algebraic, group
theory and symmetry-enforcing techniques, and that describe the
interaction between fermions and bosons. Some examples of this
class of models can be found in \cite{schutte}, \cite{geyer}, and
\cite{civit-reboi}. Concerning the use of symmetry principles, the
work of  \cite{gluons99} shows how the ordering of levels of many
gluon states may be understood. In \cite{gluons99}, the predictive
power of the model, based on the microscopical treatment of gluons
configurations, was tested  by means of the spectrum related to
exotic quantum numbers.

The purpose  of this work is to extend the model of Ref.
\cite{paperI} to describe the main structure of the hadronic spectrum of
QCD.
The essentials of the model are the following:

a)color, flavor and
spin degrees of freedom are taken explicitly into account to built
the configurations, both in the quark and gluon sectors of the
model;

b)the quarks and antiquarks are placed in a s-state;

c)the
interaction of the quarks with the gluons proceeds via gluon
pairs coupled to spin and color zero, only. Other possible gluon
states are considered as spectators.

The present work is organized as follows. Section \ref{s2} gives a brief
description of the model.
We shall avoid as much as possible the repetition of
details which can be found in the already published papers \cite{paperI,paperII, paperIII},
and concentrate in the aspects which are relevant for the description of the
hadronic spectrum.
In Section \ref{s3} we discuss the
classification of many quark-antiquark states, and the
classification of many gluon states \cite{gluons99}.
In Section \ref{s4}
we show and discuss the hadronic spectrum predicted by the model,
as a natural continuation of the study presented in \cite{paperI},
where the meson sector was analyzed. Conclusions are drawn in
Section \ref{s5}.

\section{The model}\label{s2}

The basic degrees of freedom of the model are illustrated in the
level scheme shown in Figure 1. The quarks occupy two
levels, one at negative and the other at positive energy.
Antiquarks are depicted as holes in the lower level. If with
$c^\dagger_{\alpha i}$ and $c^{\alpha i}$ we denote the creation
and annihilation operators of quarks, in the upper ($i=1$) and
lower ($i=2$) level, the relation to quark (antiquark) creation
and annihilation operators is given by
\begin{eqnarray}
\bd{a}^\dagger_{\alpha} & = & \bd{c}^\dagger_{\alpha 1}, ~~~
\bd{d}_{\alpha}  =  \bd{c}^\dagger_{\alpha 2}\nonumber \\
\bd{a}^{\alpha} & = & \bd{c}^{\alpha 1}, ~~~
\bd{d}^{\dagger ~\alpha}  =  \bd{c}^{\alpha 2}~~~,
\label{c}
\end{eqnarray}
where the index $\alpha$ refers to color, flavor and spin quantum
numbers. Since three degrees of color, three degrees of flavor (only
u, d and s quarks are taken into account) and two degrees of spin
are considered, each level is 18-fold degenerate. The energy of each
level is approximated to one third of the nucleon mass and it represents a
constituent (effective) mass of the quarks. In this picture, the mass
comes mainly from the kinetic energy because quarks are confined to a
sphere of radius 1fm. The model describes quarks and antiquarks
distributed in an orbital s-level. Orbital excitations can also be
accounted for by increasing the number of degrees of freedom.
\begin{figure}
\rotatebox{270}{\includegraphics[width=10cm,height=10cm]{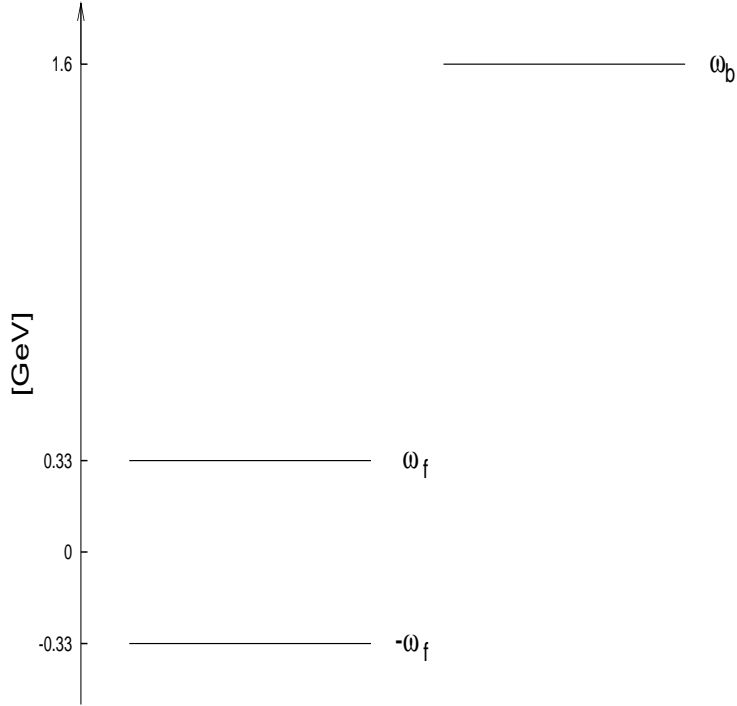}}
\vskip 0.5cm
\caption{
Schematic representation of
the model space. The fermion levels are indicated by their
energies $\pm \omega_f$. The gluon-pairs are represented by the
level at the energy $\omega_b$.
}
\label{fig1}
\end{figure}
To this system of quarks we add a boson level, which represents
pairs of gluons of color and spin zero. Its position is fixed as it was done in
Ref. \cite{gluons99}. The coupling between quarks and antiquarks
and gluons proceeds via this gluon-pair configurations. Other
couplings to different pairs of gluons are neglected, because
their contributions, at least for the meson spectrum, were found to be
negligible \cite{paperI}. Nevertheless, gluons of other types are
still present in the model but they are considered as spectators.

The basic excitations in the fermion sector are $q{\bar q}$ pairs,
which are described by $B^\dagger_{\lambda f,SM}$ and $B^{\lambda
f,SM}$ pair operators \cite{paperI}, where $\lambda$=0,1 stands
for flavor states (0,0) and (1,1),  and the spin $S$ can be 0 or
1. These pair of fermions, a sub-class of bi-fermion operators
\cite{geyer,civit-reboi}, can be mapped onto a boson space
\cite{klein}. We denote the corresponding operators by
$b^\dagger_{\lambda f,SM}$.

This is the basic structure of the model
presented previously \cite{paperI}.
In order to extend it to
the description of baryons, and because we work in the boson
space of quark-antiquark pairs, one should add three valence
quarks via the operators
$[ D^{(\lambda_0,\mu_0)S_0}\times
a^\dagger ]^{(\lambda , \mu      )}_{f,m}$, where two {\it ideal} quark
operators $a^\dagger_\alpha$ \cite{klein} are coupled as a di-quark
$D^{(\lambda_0,\mu_0)S_0}_{f_0 m_0}$ =
$[a^\dagger \times a^\dagger ]^{(\lambda_0,\mu_0)S_0}_{f_0 m_0}$
with flavor, spin $(\lambda_0,\mu_0)S_0$ either (0,1)0 or (2,0)1.

The model Hamiltonian is written as in Ref. \cite{paperI}, with
the inclusion of new terms which are needed for the description of baryons,
namely:
\begin{eqnarray}
\bd{H} & = & 2\omega_f \bd{n_f} + \omega_b \bd{n_b} + \nonumber \\
& & \sum_{\lambda S} V_{\lambda S}
\left\{ \left[ (\bd{b}_{\lambda S}^\dagger )^2 +
2\bd{b}_{\lambda S}^\dagger \bd{b}_{\lambda S} + (\bd{b}_{\lambda S})^2 \right]
(1-\frac{\bd{n_f}}{2\Omega})\bd{b} + \right.
\nonumber \\
& & \left. \bd{b}^\dagger (1-\frac{\bd{n_f}}{2\Omega})
 \left[ (\bd{b}_{\lambda S}^\dagger )^2 +
2\bd{b}_{\lambda S}^\dagger \bd{b}_{\lambda S}+ (\bd{b}_{\lambda S})^2
\right] \right\}    +   \nonumber \\
& & \bd{n}_{(0,1)0} \left( D_1 \bd{n_b} + D_2(\bd{b}^\dagger + \bd{b}) \right)
+
\bd{n}_{(2,0)1} \left( E_1 \bd{n_b} + E_2(\bd{b}^\dagger + \bd{b}) \right)
~~~,
\label{hamiltonian}
\end{eqnarray}
where $(\bd{b}_{\lambda S}^\dagger )^2$ $=$ $(\bd{b}_{\lambda
S}^\dagger \cdot \bd{b}_{\lambda S}^\dagger )$ is a short hand
notation for the scalar product \footnote{ Similarly for $(\bd{b}_{\lambda
S})^2$ and $(\bd{b}_{\lambda S}^\dagger \bd{b}_{\lambda S})$}. The
factor $(1-\frac{\bd{n_f}}{2\Omega})$ simulates the effect of
the terms which
would appear in the exact boson mapping of the quark-antiquark
pairs \cite{klein, civit-reboi}. The operators $\bd{b}^\dagger$
and $\bd{b}$ are  boson creation and
annihilation operators of the gluon pairs with spin and
color zero.

The interaction describes scattering and vacuum
fluctuation terms of fermion and gluon pairs. The strength
$V_{\lambda S}$ is the same for each allowed value of $\lambda$
and $S$, due to symmetry reasons, as shown in \cite{paperI}.
These parameters were determined in \cite{paperI} and they are
taken as fixed values in the present calculations.
The matrix
elements of the meson part of the
Hamiltonian are calculated in a seniority basis.
For details, please see Ref. \cite{paperI}.
The operators $\bd{n}_{(\lambda_0,\mu_0)S_0}$ are  number operators
of the di-quarks coupled to flavor-spin $(\lambda_0,\mu_0)S_0$.
The interaction does not contain terms which connect
states with different hypercharge and isospin. It does not
contain flavor mixing terms, either. Therefore, in the comparison with data,
the energies should be corrected by the mass formula given in
Ref. \cite{gursey}, with parameters deduced as in \cite{roelof-penta}
and turning off flavor mixing terms.
The procedure is explained in details in \cite{paperI}.

The  parameters $D_k$ are adjusted to the nucleon resonances,
while the  parameters $E_k$ are adjusted to the $\Delta$
resonances. The two last terms in (\ref{hamiltonian}) are needed
to describe the interaction between the valence quarks and the
meson-like states. The parameters $\omega_f$ and $\omega_b$ were
fixed at 0.33 GeV and 1.6 GeV, respectively. The parameters
$V_{\lambda S}$ were adjusted to the mesonic spectrum and their
values are: $V_{00}$=0.0337 GeV, $V_{01}$=0.0422 GeV,
$V_{10}$=0.1573 GeV and $V_{11}$=0.0177 GeV, respectively. This
shows the strong effects of the $V_{10}$ interaction, which acts
on pairs coupled to flavor (1,1) and spin 0 (pion-like states ).
The parameters $D_k$ and $E_k$ (k=1,2) are adjusted to the nucleon
resonances and $\Delta$ resonances respectively. Their values are
$D_1=$-1.442 GeV, $D_2$=-0.439 GeV, $E_1$=-1.187 GeV and
$E_2$=-0.362 GeV.

\section{ The model space}\label{s3}

The complete classification of the many quark-antiquark system was
given in Ref. \cite{paperI}. Here we will summarize it and discuss
the aspects which are relevant for the present calculations. The
main idea about the classification of the many quark-antiquark
system is to treat quarks in the two level system, as depicted in
Fig. 1. The ground state (vacuum), for the case of no interaction,
is given by 18 quarks occupying the lowest level. A $n$
particle-hole excitation of this ground state represents $n$
physical quarks and antiquarks. Thus, baryons are represented by
three physical quarks in the upper level. Particles with a
structure of the form $q^3(q\bar{q})^n$ are then described by
three valence quarks plus $n$ particle-hole excitations. The group
chain, describing these states, is given by
\begin{eqnarray}
[1^N] &  [h]=[h_1h_2h_3] & [h^T] \nonumber \\
U(4\Omega )  & \supset  U(\frac{\Omega}{3}) ~~~~~~~~~~~ \otimes & U(12) \nonumber \\
&  ~~~ \cup ~~~ & \cup \nonumber \\
&  ~~ (\lambda_C,\mu_C)~SU_C(3) ~~~~ [v_1v_2v_3v_4] & U(4) \otimes SU_f(3) (\lambda_f,\mu_f)
\nonumber \\
& & \cup \nonumber \\
& & SU_S(2) ~S,M ~~~,
\label{group1}
\end{eqnarray}
($\Omega = 9$ is the degeneracy corresponding to  3 degrees of
freedom in the color sector and three degrees of freedom in
flavor) \cite{paperI}, where the irreducible representation
(irrep) of $U(4\Omega )$ is the completely antisymmetric one and
$N$ is the number of particles involved. Similar group chains can
be constructed for larger values of $\Omega$. The upper index in
$[h^T]$ refers to the transposed Young diagram of $[h]$, where the
columns and rows are interchanged \cite{hamermesh}. Due to the
antisymmetric irrep $[1^N]$ of $U(4\Omega)$ the irreps of
$U(\Omega /3)$ and $U(12)$ are complementary and the irrep of
$U(\Omega /3)$, which for $\Omega =9$ is the color group, has
maximally three rows \cite{hamermesh}. In the group chain
(\ref{group1}) no multiplicity labels are indicated. There is a
multiplicity $\rho_f$ for $(\lambda_f,\mu_f)$ and $\rho_S$ for the
spin $S$. The color labels $(\lambda_C,\mu_C)$ are related to the
$h_i$ via $\lambda_C = h_1-h_2$ and $\mu_C = h_2-h_3$. The irrep
of the $U(4)$ group is given by a Young diagram with four rows,
i.e. $[v_1,v_2,v_3,v_4]$. The complete state is given by
\begin{eqnarray}
|N, [v_1v_2v_3v_4] (\lambda_C ,\mu_C), \rho_f (\lambda_f,\mu_f) Y T T_z, \rho_S S M> ~~~,
\label{state}
\end{eqnarray}
where $Y$ is the hypercharge, $T$ is the isospin and $T_z$ its
third component. For meson-like states, the color quantum numbers
to be considered are $(\lambda_C, \mu_C)=(0,0)$.

To obtain the values of $h_i$ one has to consider all possible
partitions of $N=h_1+h_2+h_3$. For
colorless states we have $h_1=h_2=h_3=h$. Each partition of $N$
appears only once. The irrep $[hhh]$ of
$U(\frac{\Omega}{3})=U(3)$ ($\Omega=9$) fixes the irrep of
$U(12)$, as indicated in (\ref{group1}).
For $\Omega = 9$ and color zero the meson irrep of $U(12)$ is given by $[3^6 0^6]$
and for baryons it is $[3^7 0^5]$.

As an example, in Table \ref{table1} we list the partial
$U(4)\times U_f(3)$ content for meson states of the $U(12)$ irrep
$[3^6 0^6]$ and in Table \ref{table2} the partial content of the
$U(4)$ irreps listed in Table \ref{table1}. The irreps which are
shown  contain up to two quark-antiquarks pairs, only\footnote{
These tables are constructed with the help of a computer code
which may be found in \cite{comp-sergio}}. The $U(4)$ irreps have
in general four rows while the $U_f(3)$ of the flavor group have
three rows $[p_1 p_2 p_3]$. The adopted notation for flavor is
$(\lambda , \mu )$ where $\lambda = p_1 - p_2$ and $\mu = p_2 -
p_3$. The minimal number of quarks in an $U(4)$ irrep is given by
the sum of the two last rows of the $U(4)$ Young diagram while the
number of antiquarks is given by the difference between  $2\Omega
= 18$ and the sum of the two first rows of the $U(4)$ Young
diagram. In the spin representation for quarks ($[q_1 q_2]$), the
sum $n_q = q_1 + q_2$ gives the number of quarks and $S_q = (q_1
-q_2)/2$ is the spin carried by the quarks. Similarly, for
antiquarks($[\bar{q}_1 \bar{q}_2]$) $S_{\bar{q}} = (\bar{q}_1
-\bar{q}_2)/2$ and $2\Omega-n_{\bar{q}} = \bar{q}_1 + \bar{q}_2$.
This is related to the fact that the unperturbed ground state, of
the two level model space, contains 18 quarks, as it was defined
before. The total spin $S$ of the system can be obtained via the
condition $|S_q-S_{\bar{q}}|\le S \le S_q+S_{\bar{q}}$. In Tables
\ref{table11} and \ref{table22} the classification of the baryon
states is partly given. In  Table \ref{table11} only $U(4)\otimes
U_f(3)$ states are considered which contain five quarks and two
antiquarks. In Table \ref{table22} only lowest states which
contain unusual flavor are listed. The $U(12)$ irrep is given by
$[3^7 0^5]$.
\begin{center}
\begin{table}
\begin{center}
\begin{tabular}{|l|l|l|l|}
\hline
$U(4)$ & $U_f(3)$ & $(\lambda , \mu )$ & mult \\
\hline
 $[ 8  8  1  1]$    &   $[ 6  6  6]$ & (0,0) &    1 \\
 $[ 8  8  1  1]$    &   $[ 7  6  5]$ & (1,1) &    1 \\
 $[ 8  8  1  1]$    &   $[ 8  6  4]$ & (2,2) &    1 \\
 $[ 9  7  1  1]$    &   $[ 7  6  5]$ & (1,1) &    1 \\
 $[ 9  7  1  1]$    &   $[ 7  7  4]$ & (0,3) &    1 \\
 $[ 9  7  1  1]$    &   $[ 8  5  5]$ & (3,0) &    1 \\
 $[ 8  8  2  0]$    &   $[ 7  6  5]$ & (1,1) &    1 \\
 $[ 8  8  2  0]$    &   $[ 7  7  4]$ & (0,3) &    1 \\
 $[ 8  8  2  0]$    &   $[ 8  5  5]$ & (3,0) &    1 \\
 $[ 9  7  2  0]$    &   $[ 6  6  6]$ & (0,0) &    1 \\
 $[ 9  7  2  0]$    &   $[ 7  6  5]$ & (1,1) &    1 \\
 $[ 9  7  2  0]$    &   $[ 8  6  4]$ & (2,2) &    1 \\
 $[ 9  8  1  0]$    &   $[ 7  6  5]$ & (1,1) &    1 \\
 $[ 9  9  0  0]$    &   $[ 6  6  6]$ & (0,0) &    1 \\
\hline
\end{tabular}
\end{center}
\caption{ Partial $U(4)\times U_f(3)$ content of the $U(12)$ irrep
$[3^60^6]$ (color zero, meson like states). The flavor irreps
(0,0), (1,1), (3,0), (0,3), (2,2) describe the flavor singlet,
octet, decuplet, anti-decuplet, and 27-plet, respectively. The
multiplicity of the $U(4)\times U_f(3)$ content in the $U(1)$
irrep $[3^60^6]$ is given in the last column. } \label{table1}
\end{table}
\end{center}
\begin{center}
\begin{table}
\begin{center}
\begin{tabular}{|l|l|l|l|l|l|l|l|}
\hline
$U(4)$ & $[q_1 q_2]$ & $n_q$ & $S_q$ & $[\bar{q}_1 \bar{q}_2]$ & $n_{\bar{q}}$
& $S_{\bar{q}}$ & $S$ \\
\hline
$[8811]$ & $[11]$ & 2 & 0 & $[88]$ & 2 & 0 & 0 \\
$[9711]$ & $[11]$ & 2 & 0 & $[97]$ & 2 & 1 & 1 \\
$[8820]$ & $[20]$ & 2 & 1 & $[88]$ & 2 & 0 & 1 \\
$[9720]$ & $[20]$ & 2 & 1 & $[97]$ & 2 & 1 & 0, 1, 2 \\
$[9810]$ & $[10]$ & 1 & $\frac{1}{2}$ & $[98]$ & 1 & $\frac{1}{2}$ & 0, 1 \\
$[9810]$ & $[11]$ & 2 & 0 & $[97]$ & 2 & 1 & 1 \\
$[9810]$ & $[11]$ & 2 & 0 & $[88]$ & 2 & 0 & 0 \\
$[9810]$ & $[20]$ & 2 & 1 & $[97]$ & 2 & 1 & 0, 1, 2 \\
$[9810]$ & $[20]$ & 2 & 1 & $[88]$ & 2 & 0 & 1 \\
$[9900]$ & $[00]$ & 0 & 0 & $[99]$ & 0 & 0 & 0 \\
$[9900]$ & $[10]$ & 1 & $\frac{1}{2}$ & $[98]$ & 1 & $\frac{1}{2}$ & 0, 1 \\
$[9900]$ & $[20]$ & 2 & 1 & $[97]$ & 2 & 1 & 0, 1, 2 \\
\hline
\end{tabular}
\end{center}
\caption{ Quark-antiquark content of some $U(4)$ meson irreps. The
table shows the irreps which contain at most two quarks and two
antiquarks, only. } \label{table2}
\end{table}
\end{center}
\begin{center}
\begin{table}
\begin{center}
\begin{tabular}{|l|l|l|l|}
\hline
$U(4)$ & $U_f(3)$ & $(\lambda , \mu )$ & mult \\
\hline
 $[ 8  8  3  2]$    &   $[ 10  7  4]$ & (3,3) &    1 \\
 $[ 8  8  4  1]$    &   $[ 10  7  4]$ & (3,3) &    1 \\
 $[ 9  7  3  2]$    &   $[ 10  7  4]$ & (3,3) &    1 \\
\hline
\hline
 $[ 9  8  2  2]$    &   $[ 7  7  7]$ & (0,0) &    1 \\
 $[ 9  8  2  2]$    &   $[ 8  7  6]$ & (1,1) &    1 \\
 $[ 9  8  2  2]$    &   $[ 9  6  6]$ & (3,0) &    1 \\
 $[ 9  8  2  2]$    &   $[ 9  7  5]$ & (1,1) &    1 \\
 $[ 9  8  3  1]$    &   $[ 7  7  7]$ & (0,0) &    1 \\
 $[ 9  8  3  1]$    &   $[ 8  7  6]$ & (1,1) &    2 \\
 $[ 9  8  3  1]$    &   $[ 9  6  6]$ & (3,0) &    1 \\
 $[ 9  8  3  1]$    &   $[ 8  8  5]$ & (0,3) &    1 \\
 $[ 9  8  3  1]$    &   $[ 9  7  5]$ & (2,2) &    1 \\
 $[ 9  8  3  1]$    &   $[ 10  6  5]$ & (4,1) &    1 \\
 $[ 8  8  5  0]$    &   $[ 7  7  7]$ & (0,0) &    1 \\
 $[ 8  8  5  0]$    &   $[ 8  7  6]$ & (1,1) &    1 \\
 $[ 8  8  5  0]$    &   $[ 8  8  5]$ & (0,3) &    1 \\
 $[ 8  8  5  0]$    &   $[ 9  7  5]$ & (2,2) &    1 \\
 $[ 9  8  4  0]$    &   $[ 8  7  6]$ & (1,1) &    1 \\
 $[ 9  8  4  0]$    &   $[ 9  7  5]$ & (2,2) &    1 \\
 $[ 9  8  2  1]$    &   $[ 8  7  6]$ & (1,1) &    1 \\
 $[ 9  9  3  0]$    &   $[ 9  6  6]$ & (3,0) &    1 \\
\hline
\end{tabular}
\end{center}
\caption{ Partial $U(4)\times U_f(3)$ content of the $U(12)$ irrep
$[3^70^5]$ (color zero, baryon like states). The flavor irreps
(0,0), (1,1), (3,0), (0,3), (2,2), (4,1) describe the flavor
singlet, octet, decuplet, anti-decuplet, 27-plet, and 35-plet,
respectively. The multiplicity of the $U(4)\times U_f(3)$ content
in the $U(1)$ irrep $[3^70^5]$ is given in the last column. The
first three rows give some higher lying $U(4)$ irreps which are
associated with a flavor irrep (3,3). The double line separates
some high lying irreps from the low lying ones. } \label{table11}
\end{table}
\end{center}
\begin{center}
\begin{table}
\begin{center}
\begin{tabular}{|l|l|l|l|l|l|l|l|}
\hline
$U(4)$ & $[q_1 q_2]$ & $n_q$ & $S_q$ & $[\bar{q}_1 \bar{q}_2]$ & $n_{\bar{q}}$
& $S_{\bar{q}}$ & $S$ \\
\hline
$[8832]$ & $[32]$ & 5 & $\frac{1}{2}$ & $[88]$ & 2 & 0 & $\frac{1}{2}$ \\
$[8841]$ & $[41]$ & 5 & $\frac{3}{2}$ & $[88]$ & 2 & 0 & $\frac{3}{2}$ \\
$[9732]$ & $[32]$ & 5 & $\frac{1}{2}$ & $[97]$ & 2 & 1 & $\frac{1}{2}$, $\frac{3}{2}$ \\
\hline
\hline
$[9831]$ & $[41]$ & 5 & $\frac{3}{2}$ & $[88]$ & 2 & 0 & $\frac{3}{2}$ \\
$[9831]$ & $[32]$ & 5 & $\frac{1}{2}$ & $[88]$ & 2 & 0 & $\frac{1}{2}$ \\
$[9831]$ & $[32]$ & 5 & $\frac{1}{2}$ & $[97]$ & 2 & 1 & $\frac{1}{2}$, $\frac{3}{2}$ \\
$[8850]$ & $[50]$ & 5 & $\frac{5}{2}$ & $[88]$ & 2 & 0 & $\frac{5}{2}$ \\
$[9840]$ & $[50]$ & 5 & $\frac{5}{2}$ & $[88]$ & 2 & 0 & $\frac{5}{2}$ \\
$[9840]$ & $[50]$ & 5 & $\frac{5}{2}$ & $[97]$ & 2 & 1 & $\frac{3}{2}$, $\frac{5}{2}$, $\frac{7}{2}$ \\
$[9840]$ & $[41]$ & 5 & $\frac{3}{2}$ & $[97]$ & 2 & 1 & $\frac{1}{2}$, $\frac{3}{2}$, $\frac{5}{2}$ \\
$[9831]$ & $[31]$ & 4 & 1 & $[98]$ & 1 & $\frac{1}{2}$ & $\frac{1}{2}$, $\frac{3}{2}$ \\
$[9840]$ & $[40]$ & 4 & 2 & $[98]$ & 1 & $\frac{1}{2}$ & $\frac{3}{2}$, $\frac{5}{2}$ \\
$[9921]$ & $[21]$ & 3 & $\frac{1}{2}$ & $[0]$ & 0 & 0 & $\frac{3}{2}$  \\
$[9930]$ & $[30]$ & 3 & $\frac{3}{2}$ & $[0]$ & 0 & 0 & $\frac{3}{2}$ \\
\hline
\end{tabular}
\end{center}
\caption{ Quark-antiquark content of some $U(4)$ baryon irreps.
Only those irreps are shown which contain either the nucleon and
$\Delta$ resonances (last two rows) or penta- and heptaquark
states with unusual flavor. There are more irreps at lower energy
but with a flavor which can also be reached by a three quark
system. The first three rows shows the content of those irreps
which are associated with flavor irreps (3,3) (see first three
rows in Table \ref{table11}). The double line separates some high
lying irreps from the low lying ones. } \label{table22}
\end{table}
\end{center}
To extend the meson-like configurations discussed in Ref.
\cite{paperI}, we shall include the flavor irreps (3,0), (0,3) and
(2,2), since all of these contain at least two quark-antiquark
pairs. The irreps (3,0) and (0,3) have total spin 1 while the
irrep (2,2) may have total spin values 0, 1, and 2. Within our
boson representation of the quark-antiquark pairs, the (3,0) and
(0,3) irreps with spin 1 can be obtained via the coupling scheme
(e.g. for (3,0)) $\left[ b^\dagger_{(1,1)0} \times
b^\dagger_{(1,1)1}\right]^{(3,0)1}_{f_0 m=1}|0>$ while the
irrep(2,2)$S=0,1,2$ comes from $\left[ b^\dagger_{(1,1)1} \times
b^\dagger_{(1,1)1}\right]^{(2,2)S=0,1,2}_{f_0 m=S}|0>$, $\left[
b^\dagger_{(1,1)0} \times
b^\dagger_{(1,1)0}\right]^{(2,2)S=0}_{f_0 m=S}|0>$. The two
configurations with $S=0$ are linear combinations of the $U(4)$
irreps [9720] and [8811]. For $S=1$ the coupling scheme is $\left[
b^\dagger_{(1,1)0} \times
b^\dagger_{(1,1)1}\right]^{(2,2)S=1}_{f_0 m=S}|0>$, where $f_0$
denotes the flavor state with maximum weight.

In the boson representation the states are given by the direct product
of the eigenstates of one-, three-, eight- and 24-dimensional harmonic
oscillators \cite{paperI}. For each harmonic oscillator
the basis states are given by

\begin{eqnarray}
{\cal N}_{N_{\lambda S}\nu_{\lambda S}} (\bd{b}^\dagger_{\lambda S}
\cdot \bd{b}^\dagger_{\lambda S})^{\frac{N_{\lambda S}-\nu_{\lambda S}}{2}}
|\nu_{\lambda S}\alpha_{\lambda S} > ~~~,
\label{basis}
\end{eqnarray}
where $N_{\lambda S}$ is the number of bosons of type
$[\lambda ,S]$, $\nu_{\lambda S}$ is the seniority
and ${\cal N}_{N_{\lambda S}\nu_{\lambda S}}$ is a normalization constant.
The seniority is the number of uncoupled $\bd{b}_{\lambda S}$-bosons.
The quantity $\alpha_{\lambda S}$ contains all other quantum numbers needed to
specify a particular harmonic oscillator.
The state (\ref{basis}) can be viewed as the superposition of a certain number of
quark-antiquark pairs coupled to a given flavor-spin combination.
This number of quark-antiquark pairs will be used
to denote non-trivial configurations.

The Hamiltonian changes only the number of trivial boson pairs
(of the type $(b^\dagger_{\lambda S} \cdot b^\dagger_{\lambda S})$),
and it leaves the seniority invariant.
The number of these boson pairs is not conserved, i.e.
a general state (\ref{basis}) contains a given seniority, which is related
to an integer number of quark-antiquark pairs, and a number of
boson pairs which is not fixed. Because the model Hamilton only
changes the number of boson pairs, we can divide the Hilbert space
into subspaces of given combinations of seniorities of the different types
of quark-antiquark pairs with a given flavor-spin combination.
The spectrum obtained for a given flavor-spin combination and a given
seniority combination will be identical to another one with
a different spin-flavor combination but with the same
seniority distribution. This leads to the appearance of a degeneration
in the spectrum.

A similar classification can be given for the baryon states.
Within the boson representation, the ideal valence quarks
are added to describe baryons. One has to couple the
meson-like states with the states $(1,1)\frac{1}{2}$ or with $(3,0)\frac{3}{2}$,
to which the three valence quarks can be coupled, to describe
nucleon and $\Delta$ resonances, respectively. This implies a
degeneration of different flavor-spin irreps. As an example, if the configuration
$(1,1)\frac{1}{2}^+$
of the three valence quarks is coupled to a meson irrep $(1,1)0^-$ then
the resulting allowed flavor-spin irreps are $(0,0)\frac{1}{2}^-$,
$(1,1)\frac{1}{2}^-$, $(3,0)\frac{1}{2}^-$, $(0,3)\frac{1}{2}^-$
and $(2,2)\frac{1}{2}^-$.

Another ingredient of the model is the non-conservation
of the particle number (quarks, antiquarks and gluons). This is a fundamental
property of a relativistic theory with large interaction constants. It means that
physical particles, in general, do not contain a fixed number of quarks, antiquarks
and/or gluons but rather an average number of them. In this picture,
the nucleons will not contain, mainly, three valence quarks but also an
average number of quark-antiquark and gluon pairs. The mixing
of particle number turned out to be essential in order to remove the multiplicity
of states at low energy, a feature which any constituent quark model exhibits when the number
of quarks, antiquarks and gluons is fixed.

Herewith we summarize, for completeness, the configurations which
may appear in the gluon sector of the theory. In the model
\cite{gluons99} the basic building blocks are effective gluons
with color (1,1) and spin 1. The detailed group structure and the
model Hamiltonian used is discussed in Ref. \cite{gluons99}. In
Table \ref{glue} we show all states of glue-balls with color zero
and maximally six effective gluons. The lowest glue-ball state is
the first excited $J^{\pi C}$ = $0^{++}$ state, where $J$ is the
angular momentum, $\pi$ is the parity and $C$ is  the charge
conjugation. For details, see Ref. \cite{gluons99}.
\begin{center}
\begin{table}
\begin{center}
\begin{tabular}{|l|l|l|l|l|}
\hline
$U(8)$ ($U(3)$) & $O(8)$ & $SO(3)$ ($J$) & $P$ & $C$ \\
\hline
$[2]$ & (0000) &  0,2 & +1 & +1 \\
$[4]$ & (0000) &  0,2,4 & +1 & +1 \\
$[2^2]$ & (0000) & 0,2 & +1 & +1 \\
$[6]$ & (0000) & 0,2,4,6 & +1 & +1 \\
$[42]$ & (0000) & 0,2$_2$,3,4 & +1 & +1 \\
$[2^3]$ & (0000) & 0 & +1 & +1 \\
$[3]$ & (3000) & 1,3 & -1 & -1 \\
$[5]$ & (3000) & 1,3,5 & -1 & -1 \\
$[41]$ & (3000) & 1,2,3,4 & -1 & -1 \\
$[32]$ & (3000) & 1,2,3 & -1 & -1 \\
$[1^3]$ & (1110) & 0 & -1 & +1 \\
$[31^2]$ & (1110) & 0,2 & -1 & +1 \\
$[21^2]$ & (2110) & 1 & +1 & -1 \\
$[41^2]$ & (2110) & 1,3 & +1 & -1 \\
$[321]$ & (2110) & 1,2 & +1 & -1 \\
$[2^2]$ & (2200) & 0,2 & +1 & +1 \\
$[42]$ & (2200) & 0,2$_2$,3,4 & +1 & +1 \\
$[321]$ & (2200) & 1,2 & +1 & +1 \\
$[2^3]$ & (2200) & 0 & +1 & +1 \\
$[2^21]$ & (2210) & 1 & -1 & -1 \\
$[31^2]$ & (3110) & 0,2 & -1 & -1 \\
$[6]$ & (6000) & 0,2,4,6 & +1 & +1 \\
$[42]$ & (4200) & 0,2$_2$,3,4 & +1 & +1 \\
$[41^2]$ & (4110) & 1,3 & +1 & -1 \\
$[3^2]$ & (3300) & 1,3 & +1 & -1 \\
$[2^3]$ & (2220) & 0 & +1 & +1 \\
\hline
\end{tabular}
\end{center}
\caption{ The classification of many-gluon states, with color
(0,0). The multiplicity of the configuration (0,0) is one up to
six gluons. For the notation of the groups, which appear in the
many-gluons states, and their irreps see Ref. \cite{gluons99},
where the general classification, i.e. with open color, can be
deduced. $P$ and $C$ refer to the parity and charge conjugation of
the gluon states respectively. } \label{glue}
\end{table}
\end{center}

\section{Applications}\label{s4}

In this section we shall show and discuss the results of the model introduced in the previous
section, particularly, concerning baryon states.
Since the Hamiltonian conserves flavor, spin and parity, all states
belonging to the same flavor irrep are degenerate. In order to introduce
a splitting in the isospin and hypercharge the Gel'man-Okubo interaction
\cite{symm} have to be added.
Otherwise, the parameters of the Hamiltonian should be adjusted
to reproduce the corrected experimental masses.
This is the procedure which we have adopted in the
calculations.
Next Subsection \ref{s4.1} is devoted to the results concerning mesonic states, others
than the ones reported in \cite{paperI},
like those with flavor (2,2). These states illustrate the role of
some additional degeneracies of the model, and
their possible physical implications. In Subsection \ref{s4.2} we present the results
for nucleon resonances, as well as some results about more exotic states,
like the predicted  energy and parity of the lowest
penta- and hepta-quarks states. Finally, in Subsection \ref{s4.3} we analyze the case of
$\Delta$ resonances. The experimental data are taken
from Ref. \cite{databook}.

\subsection{Meson States}\label{s4.1}

In Table \ref{table4} we list the quark-antiquark and gluon contents of some selected
mesonic states. The
quantity $n_{\lambda S}$ is the average number of $q\bar{q}$ pairs
coupled to flavor $(\lambda , \mu )$ and spin $S$. The quantity $n_g$ is
the average number of gluon pairs coupled to color singlet and spin zero.
The seniority content is given by the value
$(v)$ = $(v_{00},v_{01},v_{10},v_{11})$, where
$v_{\lambda S}$ refers to the seniority in the channel
$(\lambda ,\lambda )S$.
Some of the states shown in this table belongs to
flavor irreps which contain exotic combinations of hypercharge and
isospin which are not allowed in the simplest $q\bar{q}$ system. From
the seniority content one sees that these states contain at least two
$q\bar{q}$ pairs.
\begin{center}
\begin{table}
\begin{center}
\begin{tabular}{|l|l|l|l|l|l|l|l|l|l|}
\hline particle & $E_{theo}$ & $(\lambda_f,\mu_f)$ & $J^\pi$ &
($v$) & $n_{00}$ & $n_{01}$
& $n_{10}$ & $n_{11}$ & $n_g$   \\
\hline
vacuum & 0.0    & (0,0) & $0^+$ & (0000) & 0.00 & 0.03  & 3.11 & 0.03 & 1.705 \\
$f_0$(400-1200)       & 0.656  & (0,0) & $0^+$ & (0000)  & 0.00 & 0.01 & 0.46 & 0.01 & 0.32 \\
$f_0$(980)       & 0.797  & (1,1) & $0^+$ & (0020) & 0.00 & 0.02 & 3.78 & 0.02 & 1.49 \\
$f_1$(1420)       & 1.445  & (0,0) & $1^+$ & (0011) & 0.00 & 0.02 & 2.39 & 1.02 & 0.90  \\
$f_2$(1270)       & 1.363  & (1,1) & $1^+$ & (0110)  & 0.00 & 1.03 & 2.46 & 0.02 & 0.99  \\
*$\eta^\prime$(958)* & 0.885 & (0,0) & $0^-$ & (1000) & 1.01 & 0.02 & 2.51 & 0.02 & 1.29  \\
$\eta$(1440)            & 1.379   & (0,0) & $0^-$ & (1000)  & 1.00 & 0.01 & 0.77 & 0.01 & 0.44  \\
*$\eta$(541)* & 0.602       & (1,1) & $0^-$ & (0010)  & 0.00 & 0.02 & 2.71 & 0.03 & 1.16 \\
*$\eta$(1295)*             & 1.428  & (1,1) & $0^-$ & (0010) & 0.00 & 0.01 & 1.61 & 0.01 & 0.53  \\
$\eta$(1760)             & 1.671  & (1,1) & $0^-$ & (1020)  & 1.01 & 0.02 & 3.53 & 0.02 & 1.25 \\
*$\omega$(782)*     & 0.851  & (0,0) & $1^-$ & (0100) & 0.04 & 1.03 & 2.56 & 0.02 & 1.34 \\
*$\phi$(1020)*       & 0.943  & (1,1) & $1^-$ & (0001) & 0.00 & 0.02 & 2.39 & 1.02 & 1.19 \\
$\omega$(1420)     & 1.389  & (1,1) & $1^-$ & (0001) & 0.00 & 0.01 & 0.85 & 1.01 & 0.47 \\
$\omega$(1600)     & 1.639  & (1,1) & $1^-$ & (0120) & 0.00 & 1.03 & 3.55 & 0.02 & 1.28 \\
$X$(1440)     & 1.440  & (3,0) & $1^+$ & (0011) & 0.00 & 0.02 & 2.39 & 1.02 & 0.90 \\
$X$(1440)     & 1.440  & (0,3) & $1^+$ & (0011) & 0.00 & 0.02 & 2.39 & 1.02 & 0.90 \\
$X$(797)     & 0.797  & (2,2) & $0^+$ & (0020) & 0.00 & 0.02 & 3.78 & 0.02 & 1.49 \\
\hline
\end{tabular}
\end{center}
\caption{ Particle content for selected meson states. In the
columns we indicate the theoretical energy $E_{theo}$, the flavor
$(\lambda_f,\mu_f)$, spin $J$ and parity $\pi$, seniority content
($v$) = $(v_{00}v_{01}v_{10}v_{11})$, expectation value of the
boson pairs in the channels (0,0) $0^-$ ($n_{00}$), (0,0) $1^-$
($n_{01}$), (1,1) $0^-$ ($n_{10}$) and (1,1) $1^-$ ($n_{11}$) and
the total number of gluon pairs ($n_g$) with spin 0. In the last
three rows some states, not reported in \cite{paperI}, are listed
which contain exotic combinations of hypercharge and isospin. The
experimental data are taken from \cite{databook}. The $X$(797) is
interpreted as a molecular state. Note that, for the particles in
the first (0,0), (1,1) $0^-$ and (0,0), (1,1) $1^-$ irreps, we are
listing the value of the masses without flavor mixing (they are
marked by an asterisk, see also Ref. \cite{paperI} for further
explanation). } \label{table4}
\end{table}
\end{center}
While the states with flavor (3,0) and (0,3) have energies of the
order of 1.5 GeV, the one with flavor (2,2) has a  very low
energy, nearly the same energy of the $f_0(980)$ state.
Experimentally, the $f_0$ state lies at approximately twice the
energy of the $\eta(541)$ state. Their dominant decay into
$K^+K^-$ leads to the interpretation of these states as
$K^+K^-$-molecules. Within our model the $(2,2)0^+$ state is built
upon the seniority $\nu=2$ state $[b^\dagger_{(1,1)0} \times
b^\dagger_{(1,1)0}]^{(2,2)0}_{f,0}|0>$. The highest weight state
is given by $b^\dagger_{(1,1)f_0,0} b^\dagger_{(1,1)f_0,0}|0>$,
which is nothing but the direct product of two pairs of the type
(1,1)0$^-$. This can be interpreted, within the present model, as
a configuration of two $q\bar{q}$-like mesons. The difference
between the mass of the two pairs and the mass of the $(2,2)0^+$
state is about 0.4 GeV, which is larger than the observed value.
One has to keep in mind that the present model is rather schematic
and that it does not include a flavor mixing interaction which
could mix $(1,1)0^-$ states with $(2,2)0^-$ states, and may induce
a coupling from, e.g., the $f_0$ state to molecular states. The
model predicts also states of the type $(q\bar{q})^2$ which can be
coupled to flavor irreps (3,0) and (0,3) (see Tables \ref{table1}
and \ref{table2}). The distribution of seniorities is (0011) and
the states lie at about 1.44 GeV. The irreps (3,0) and (0,3)
contain isospin-hypercharge configurations which cannot be reached
by a single quark-antiquark pair.

\subsection{Nuclear resonances, penta- and hepta-quarks}\label{s4.2}

In Table \ref{table5} we show the results for some selected
nucleon resonances, as they appear in the schematic model. The
content of $q\bar{q}$ pairs of the type (0,0)0$^-$ and (0,0)1$^-$
are not listed because they are always small for the cases
considered. Like for the case of mesonic states, the Hamiltonian
commutes with spin, parity, color and flavor. All states belonging
to the same flavor irrep are degenerate. In order to remove the
degeneracy one may still add the Gel'man-Okubo term plus a term
describing the flavor dependence. We follow the prescription of
Ref. \cite{gursey} with the parameters given in
\cite{roelof-penta}, instead.
\begin{center}
\begin{table}
\begin{center}
\begin{tabular}{|l|l|l|l|l|l|l|l|l|}
\hline particle & $E_{theo}$ & $(\lambda_m,\mu_m)$ & $J_m^\pi$ &
$(\lambda , \mu )J^\pi$ & ($v$) &
$n_{10}$ & $n_{11}$ & $n_g$   \\
\hline
nucleon & 0.950 & (0,0) & $0^+$ & (1,1)$\frac{1}{2}^+$ & (0000) & 0.47 & 0.05 & 1.40 \\
Roper   & 1.49  & (0,0) & $0^+$ & (1,1)$\frac{1}{2}^+$ & (0000) & 1.91 & 0.04 & 1.92  \\
N(1650)  & 1.51  & (1,1) & $0^-$ & (1,1)$\frac{1}{2}^-$ & (0010) & 3.14 & 0.03 & 2.81 \\
N(1535)  & 1.79  & (1,1) & $1^-$ & (1,1)$\frac{1}{2}^-$, $\frac{3}{2}^-$ & (0001) & 0.38 & 1.00 & 1.04 \\
$\Theta^+$(1540) & 1.51  & (1,1) & $0^-$ & (0,3)$\frac{1}{2}^-$ & (0010) & 3.14 & 0.03 & 2.81 \\
$X$(1510)  & 1.51  & (1,1) & $0^-$ & (2,2)$\frac{1}{2}^-$ & (0010) & 3.14 & 0.03 & 2.81 \\
$H_1$(2451)  & 2.45  & (1,1) & $1^+$ & (0,0), (1,1) & (0110) & 2.87 & 0.03 & 2.27 \\
 & & & & (3,0), (0,3) & & & & \\
 & & & & $\frac{1}{2}^+$, $\frac{3}{2}^+$ & & & & \\
$H_2$(2529)  & 2.53  & (2,2) & $0^+$ & (4,1), $\frac{1}{2}^+$ & (0020) & 2.73 & 0.03 & 1.24 \\
\hline
\end{tabular}
\end{center}
\caption{ Particle content for selected baryon states. In columns we
indicate the theoretical energy $E_{theo}$, the flavor
$(\lambda_m,\mu_m)$, spin $J_m$ and parity $\pi$ for the
meson part, the final flavor irrep $(\lambda , \mu )$ and spin
$J$ and parity, the seniority content ($v$) =
$(v_{00}v_{01}v_{10}v_{11})$, expectation value of the boson
pairs in the channels (1,1) $0^-$ ($n_{10}$) and (1,1) $1^-$
($n_{11}$) and the total number of gluon pairs ($n_g$) with
spin 0. In the last four rows some additional particles are
listed which contain unusual combinations of hypercharge and
isospin. The $\Theta^+$(1540) is the reported pentaquarks state.
Another pentaquarks state, according to our notation, is called
$X$ while $H_k$ ($k$=1,2) refers to the lowest heptaquarks states
with unusual flavor.
}
\label{table5}
\end{table}
\end{center}
The quark, antiquark and gluon contents of the nucleon are given
in the first row of Table \ref{table5}. The number of $q\bar{q}$
pairs is about 0.5, i.e. including the three valence quarks there
are in average 3.5 quarks and 0.5 antiquarks in the nucleon. The
number of gluon pairs is approximately 1.4, i.e. in average there
are nearly three gluons present. This implies a 59$\%$ quark
content and a 41$\%$ gluon content, a result which is expected
from previous evidences. The theoretical Roper resonance comes
near to the experimental energy of 1.44 GeV. This is a nice
result, which may be difficult to obtain in other models, except
in the constituent quark model of Ref. \cite{bijker-r}. In average
there are two $q\bar{q}$ pairs and 1.9 gluon pairs. Thus, the
Roper resonance, in the present model, contains an average number
of 5 quarks, 2 antiquarks ($q^3(q\bar{q})^2$) and 3.8 gluons. This
implies a 65$\%$ content of quarks and a 35$\%$ content of gluons.
The quark-gluon content of the Roper resonance differs from that
of the nucleons. It has more particles, a fact which is reflected
in its larger collective nature and low energy. The first negative
parity resonance appears at about 1.5 GeV, with
$(1,1)\frac{1}{2}^-$ and the next ones at 1.79GeV, with degenerate
states with $J^\pi$ = $\frac{1}{2}^-$ and $\frac{3}{2}^-$. The
first state contains a large amount of $q\bar{q}$ pairs of the
type $(1,1)0^-$ while the last states are nearly pure one
$q\bar{q}g^2$ states. These would be good candidates for hybrid
baryons. In Table \ref{table5}, states with flavor (0,3) and (2,2)
are also listed. In fact, these are the same states, concerning
their meson content. They are coupled with the three ideal valence
quarks, leading to unusual combinations of hypercharge and
isospin. These configurations may be associated to pentaquarks
states, like $\Theta^+(1540)$ \cite{penta-ex}, whose position have
been predicted in \cite{penta-pre}. Other states of the
pentaquarks type may exist, which have in common the same quantum
numbers in hypercharge and isospin as other nucleon resonances,
and, therefore, may be difficult to identify experimentally. The
interpretation of these states as pentaquarks is based on the
value of the seniority of the $q\bar{q}$ pair of the type
$(1,1)0^-$ ($\nu=1$) to which the three ideal valence quarks have
to be added. In \cite{paperIII} we present a compilation of
results about pentaquarks states, as they are predicted by our
model. Within the model, the lowest pentaquarks has negative
parity in accordance with QCD sum-rules and lattice gauge
calculations \cite{sum1,sum2,lat1,lat2}. If the orbital spin $L$
is included, pentaquarks states with positive parity may exist
with $L$=1. However, these states include an orbital excitation
and should have higher energies.

The model contains also heptaquarks, which are characterized by
two $q\bar{q}$ pairs added to the three ideal valence quarks. The
lowest one with unusual flavor, which cannot be described by a
plain three-quark system, is at about 2.5 GeV ($H_1(2451)$) (see
Table \ref{table5}). It has a  content of 2.87 $q\bar{q}$ pairs of
the type $(1,1)0^-$ ($n_{10}$)coupled with three ideal valence
quarks to the unusual flavor irrep of the type (3,0) with
spin-parity $\frac{1}{2}^+$. This implies a quark content of
78$\%$ and a gluon content of 22$\%$. There are several other
heptaquarks states near the same energy, or at slightly higher
energies, which contain one $q\bar{q}$ pair of the type $(1,1)0^-$
and one of the type $(1,1)1^-$. Also their gluon content is higher
by about one extra gluon. Within our model, the parity of the
heptaquarks state is positive. Because we are working in a boson
space, we should be careful  about the appearance of unphysical
states \cite{klein}. For example, the three valence quarks can be
coupled with the meson background with flavor (2,2), e.g., to the
flavor irreps (3,3), (4,1) and (1,4). However, only the (4,1)
irrep appears in the list of allowed states related to low lying
$U(4)$ irreps (see Tables \ref{table11} and \ref{table22}). The
others are, therefore, unphysical. There appear (3,3) irreps at
higher configurations of $U(4)$ irreps, though, with spins
$S=\frac{1}{2}$ and $\frac{3}{2}$. This shows the importance of
the complete classification, because only a comparison of the
states in the boson space with the list of irreps in the fermion
space gives the possible allowed states.

\subsection{$\Delta$ Resonances}\label{s4.3}

In a similar manner, as it was discussed in the section of nucleon
resonances, we can treat $\Delta$ resonances. In Table
\ref{table6} we list the results of our calculations.
\begin{center}
\begin{table}
\begin{center}
\begin{tabular}{|l|l|l|l|l|l|l|l|l|}
\hline particle & $E_{theo}$ & $(\lambda_m,\mu_m)$ & $J_m^\pi$ &
$(\lambda , \mu )J^\pi$ &
($v$) &
$n_{10}$ & $n_{11}$ & $n_g$   \\
\hline
$\Delta$(1232) & 1.248 & (0,0) & $0^+$ & (3,0)$\frac{3}{2}^+$ & (0000) & 0.33 & 0.03 & 0.67 \\
$\Delta$(1600)   & 1.57  & (0,0) & $0^+$ & (3,0)$\frac{3}{2}^+$ & (0000) & 1.93 & 0.04 & 1.60  \\
$\Delta$(1620)*   & 1.51  & (1,1) & $0^-$ & (3,0)$\frac{1}{2}^-$  & (0010) & 3.14 & 0.03 & 2.81  \\
$\Delta$(1700)*  & 1.79  & (1,1) & $1^-$ & (3,0), $\frac{1}{2}^-$, $\frac{3}{2}^-$ & (0001) & 0.38 & 1.00 & 1.04 \\
$\Delta$(1750)*  & 2.49  & (1,1) & $0^+$ & (3,0)$\frac{1}{2}^+$ & (1010) & 2.85 & 0.03 & 2.19 \\
$\Delta$(1900)*  & 2.09  & (1,1) & $0^-$ & (3,0)$\frac{1}{2}^-$ & (0010) & 1.40 & 0.02 & 0.68 \\
$X$(1640)  & 1.64  & (1,1) & $0^-$ & (4,1), (2,2), $\frac{3}{2}^-$ & (0010) & 3.00 & 0.03 & 2.18 \\
$H$(2530)  & 2.53  & (2,2) & $0^+$ & (4,1)$\frac{3}{2}^+$ & (0020) & 2.67 & 0.02 & 0.97 \\
\hline
\end{tabular}
\end{center}
\caption{ Particle content for selected $\Delta$ states. In columns we
indicate the theoretical energy $E_{theo}$, the flavor
$(\lambda_m,\mu_m)$, spin $J_m$ and parity $\pi$ for the
meson part, the final flavor irrep $(\lambda , \mu )$ and spin
$J$ and parity, the seniority content
$(v_{00}v_{01}v_{10}v_{11})$, expectation value of the boson
pairs in the channels (1,1) $0^-$
($n_{10}$) and (1,1) $1^-$ ($n_{11}$) and the total number of
gluon pairs ($n_g$) with spin 0. In the last two rows some
additional particles are listed
which contain unusual combinations of hypercharge and isospin.
The $X$ and $H$ states are predicted particles with
exotic hypercharge-isospin combinations. The state in the last row can be interpreted
as a heptaquarks state. The asterix indicates that the particular
state is obtained via a combination with three valence quarks
coupled to $(1,1)\frac{1}{2}^+$.
}
\label{table6}
\end{table}
\end{center}
The simplest form to obtain a $\Delta$ resonance is to couple a
diquarks with (3,0)$\frac{3}{2}$ with the meson part to (3,0)$J$,
where in general more than one $J$ is allowed, depending on the
meson state. Coupling $(3,0)\frac{3}{2}$ with $(0,0)0^+$ leads to
the $\Delta (1232)$ resonance. The quark-antiquark and gluon
content of the $\Delta (1232)$ is somewhat lower than that of the
nucleon. The theoretical state at 1.57 GeV can be compared in
structure to the Roper resonance in the nucleon resonances. The
first negative parity state with flavor (3,0) and spin
$\frac{3}{2}$ (there are other states in the table, also with
flavor (3,0) but spin $\frac{1}{2}$) is at the theoretical energy
1.79 GeV and  it contains nearly one $q\bar{q}$ pair with
flavor-spin $(1,1)1^-$.

\begin{figure}
\includegraphics[width=10cm,height=10cm]{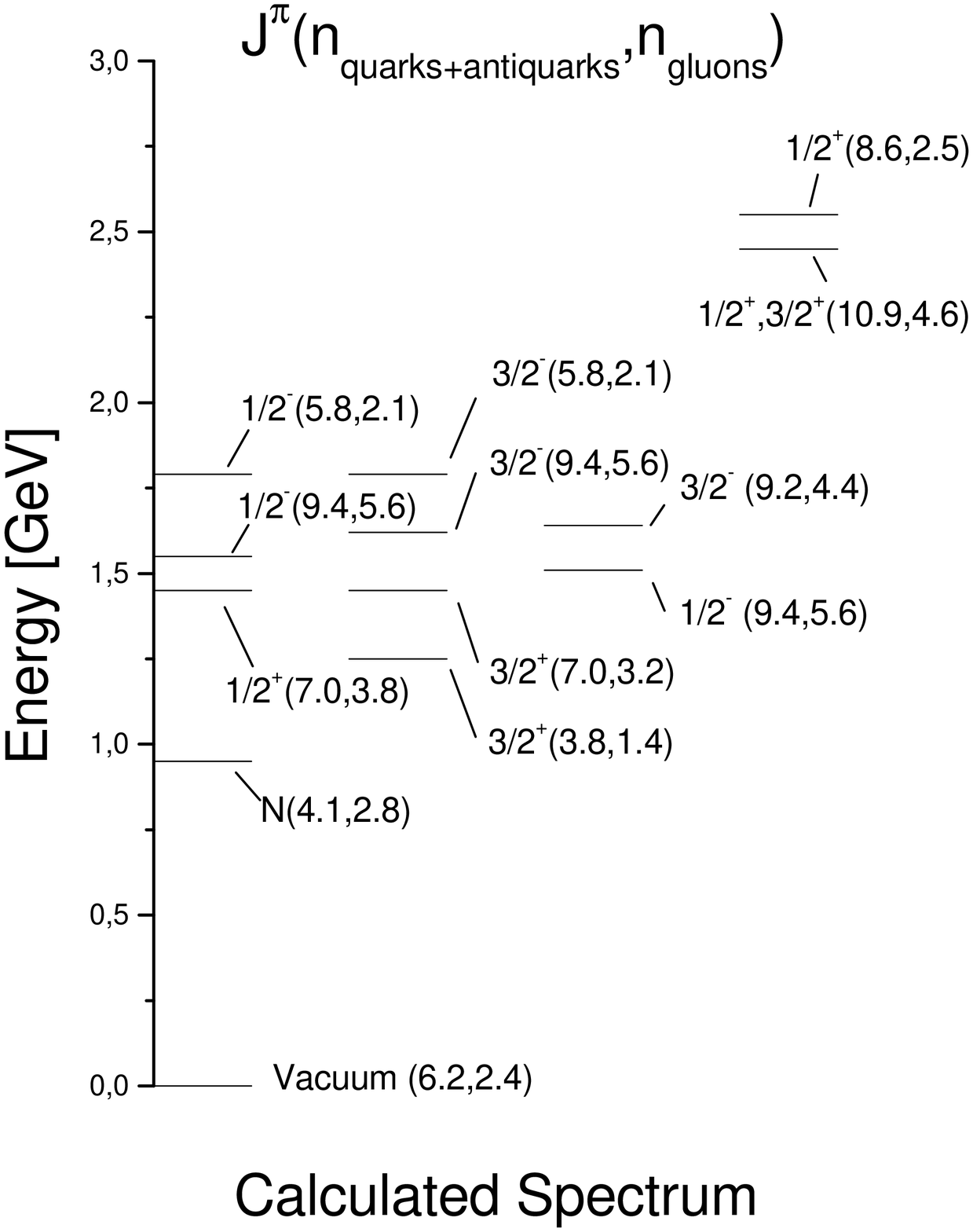}
\caption{ Nucleon resonances (first group of levels), $\Delta$
resonances (second group), pentaquarks (third group) and
heptaquarks (fourth group). On the right side of each level are
given the assigned spin and parity ($J^\pi$), and the total quark
and antiquark
($n_q+n_{\bar{q}}$) and  gluon ($n_g$) contents (see the text) The
list is not complete. Only some selected  states, of physical
interest, are listed. } \label{fig2}
\end{figure}

To conclude with this Section, we show in Figure \ref{fig2} a compilation
of the results commented upon previously.

\section{Conclusions}\label{s5}
We have extended the schematic model of \cite{paperI} to the description
of meson states, nucleon resonances and $\Delta$ resonances.
The schematic model does not conserve the number of quarks,
antiquarks and gluons, rather an average number of them can be extracted
from the calculations.
It reflects the fact that the interaction is strong and the particle
number is not conserved. It is found that
the nucleon has nearly 50$\%$ of gluon content, as expected.
The
degeneracy of states of the type $q^{n_1}\bar{q}^{n_2}g^{n_3}$
is remove (see \cite{paperI}), but there is still
a degeneracy  related to the seniority. The Hilbert space is
divided into subspaces with a given seniority of the different
$q\bar{q}$ pairs involved. Spaces with the same distribution of
seniority but different flavor and spin are degenerate. This lead
to the prediction of unusual flavor-spin combinations which
cannot be obtained by one $q\bar{q}$ pair or three quarks alone.
We gave a complete classification for the quarks, antiquarks
and gluons states. The quarks interact with pairs of gluons
coupled to spin zero. Other types of gluons are present as spectators \cite{gluons99}
The global agreement with data is reasonable and the model gives an
insight on the detailed structure of the states.
The lowest pentaquarks state is predicted  at 1.51 GeV \cite{paperIII}. It has an
unusual combination of flavor $(0,3)\frac{1}{2}^-$ and
$(2,2)\frac{1}{2}^-$, and it has negative parity. Also heptaquarks
states are predicted, with positive parity at an energy of about
2.5 GeV, for the lowest lying state.
The same was done for $\Delta$ resonances.

These are features which cannot be obtained by working alone with
three quarks  or with one quark-antiquark pair. We think that the
present results support the notion that the degrees of freedom
included in the model may indeed be the relevant ones to describe
the low-energy spectrum of QCD, within the limitations posed by
the schematic nature of the proposed Hamiltonian.

\section*{Acknowledgments}
This work belongs to the DGAPA project IN119002. It was partially supported
by the National Research Councils of Mexico (CONACYT) and Argentina
(CONICET), and by the German Academic Interchange Service of Germany
(DAAD).

\end{document}